# Toward CL-20 crystalline covalent solids: On the dependence of energy and electronic properties on the effective size of CL-20 chains

Konstantin P. Katin[a,b] and Mikhail M. Maslov[a,b,*]

[a] Department of Condensed Matter Physics, National Research Nuclear University MEPhI (Moscow Engineering Physics Institute), Kashirskoe sh. 31, 115409 Moscow, Russia

[b] Laboratory of Computational Design of Nanostructures, Nanodevices and Nanotechnologies, Research Institute for the Development of Scientific and Educational Potential of Youth, Aviatorov str. 14/55, 11960 Moscow, Russia

**Abstract.** One-dimensional CL-20 chains have been constructed using $CH_2$ molecular bridges for the covalent bonding between isolated CL-20 fragments. The energy and electronic properties of the nanostructures obtained have been analyzed by means of density functional theory and nonorthogonal tight-binding model considering Landauer-Büttiker formalism. It has been found that such systems become more thermodynamically stable as the efficient length of the chain increases. Thus, the formation of bulk covalent CL-20 solids may be energetically favorable, and such structures may possess high kinetic stability comparing to the CL-20 molecular crystals. As for electronic properties of pure CL-20 chains, they are wide-bandgap semiconductors with energy gaps equal to several electron volts that makes their use in nanoelectronic applications problematic without any additional modification.

* Corresponding author. E-mail: Mike.Maslov@gmail.com (M.M. Maslov)

## 1. Introduction

Since the first synthesis in 1987, [1] CL-20 compound ($C_6H_6N_{12}O_{12}$, hexanitrohexaazaisowurtzitane) was a subject of intensive research. It belongs to the class of high-energy-density materials (HEDM) and is of significant theoretical and practical interest. The technology for CL-20 fabrication was developed at the end of the last century [1] and, presently, the possibility of CL-20 application as the component of novel energy materials and fuel elements is considered [2]. CL-20 has one of the most unusual structures among the nitrogen-carbon cage compounds. It is formed by the strained carbon-nitrogen framework consisting of two five-membered and one six-membered rings connected by C–C bonds and six $NO_2$ functional groups attached. The molecular structure of isolated CL-20 is illustrated in Fig. 1a.

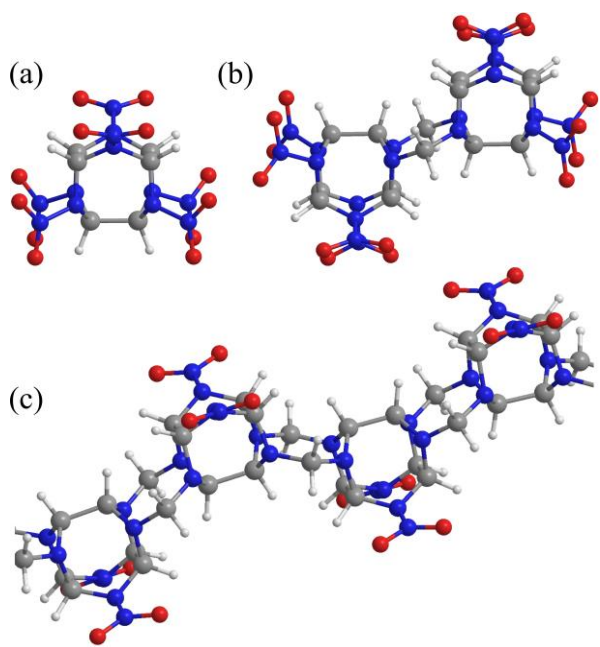

**Figure 1.** Images of isolated CL-20 molecule (a), CL-20 dimer (b), and fragment of CL-20 one-dimensional chain with the $CH_2$ molecular bridges (c)

It is well known that CL-20 exists in five different polymorphic modifications of molecular crystals (labeled as α, β, γ, ε, and ζ), differing in their molecular packing and mutual orientation of nitro groups [3-4]. However, only four of them (α, β, γ, ε) are stable under ambient conditions [5], from which the ε-modification is the most thermodynamically stable [6]. Despite the active study of both isolated CL-20 molecules and molecular solids based on their basis [1,3,7-9], there are no published data concerning the experimental synthesis or theoretical studies of CL-20 covalent crystals (at least, we failed to find them).

On the other hand, it is known that other "strained" molecules can form not only molecular complexes but covalent ones as well. For example, hydrocarbon cubane $C_8H_8$ [10], in which the carbon atoms are in cube vertices, can form both van der Waals crystal s-$C_8H_8$ [11] and linear or two-dimensional structures in which cubane fragments are connected by strong covalent bonds [12-13]. They can even form a supercubane bulk covalent structure [14]. Apparently, CL-20 molecules are capable of forming not only molecular crystals but covalent crystals as well. Recently, on the example of molecular dimers (see Fig. 1b) and tetramers we have shown the possibility of the formation of covalent bonds between the isolated CL-20 structures through different molecular bridges, which suggests the existence of covalent crystals on their basis [15]. In the present work, we go further and construct one-dimensional CL-20 chains (see Fig. 1c), analyze their energy and electronic characteristics, and estimate their thermodynamic stability. It should be noted that apart from prospects for possible practical use, studies of linear oligomers based on CL-20 fragments are of fundamental interest. The numerical simulation of these compounds is the first step toward analyzing the stability and properties of macroscopic covalent crystals on the basis of CL-20 molecules.

**2. Methods of calculation and computational details**

In our study we use density-functional theory (DFT) as well as nonorthogonal tight-binding model (NTBM). Density functional theory with B3LYP [16-17] and PBE [18] functionals with the 6-311G(d,p) electron basis set [19] is used. All DFT calculations are performed using the TeraChem program package [20-23]. Tight-binding approach is represented by originally developed for H–C–N–O systems NTBM model [24]. The parameters' fitting of NTBM is based on the criterion of the best correspondence between the computed and experimental values of binding energies, bond lengths and valence angles of several selected small $H_kC_lN_mO_n$ molecules (for parameterization details see Refs. 24-25). Previously we successfully used this tight-binding approach for electronic properties calculation of chemically functionalized cubane-based chains [26]. In general, tight-binding methods are commonly used for electronic structure calculations [27-30], including HOMO-LUMO gaps, density of states, transmission coefficients, etc.

To obtain the equilibrium structures of the CL-20 chains, we used the method of structural relaxation so that the corresponding initial configuration relaxed to a state with the local or global energy minimum under the influence of intramolecular forces only. First of all, the forces acting on the all atoms were calculated using the Hellmann-Feynman theorem. Next, atoms were shifted in the direction of the forces obtained proportional to the corresponding forces. Then the relaxation step was repeated. The whole structure is relaxed up to residual atomic forces smaller than $10^{-4}$ eV/Å. Note that for all CL-20 chains studied we also calculated the frequency spectra in the framework of the same level of theory after the geometry optimization. For each metastable configuration, the presence of local minima of energy was confirmed by the reality of all frequencies.

To calculate the electronic density of states and the conductance of infinite CL-20 chains we have applied the Landauer-Büttiker formalism [31-34] using NTBM model. The general illustration of our scheme is presented in Fig. 2.

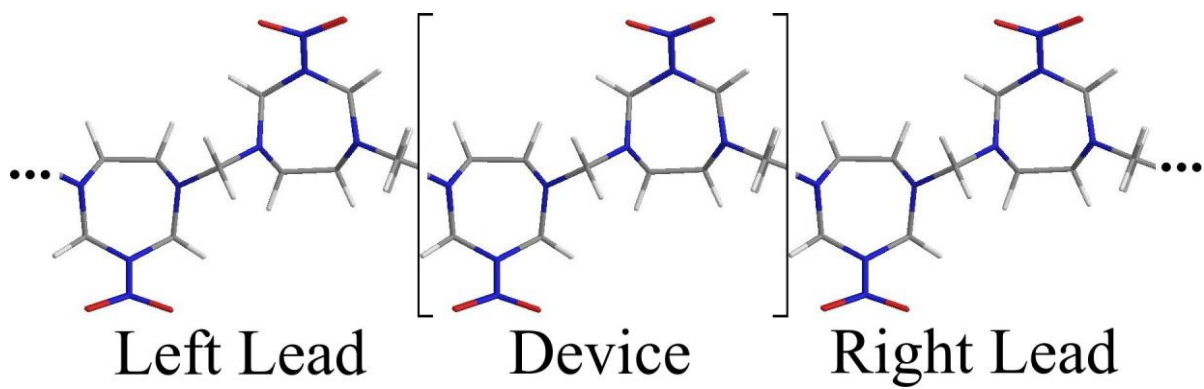

**Figure 2.** General scheme of the considered samples comprising three parts: semi-infinite left and right leads and sandwiched between them device area consisting of two CL-20 fragments

The device region (D) consisting of two CL-20 units is connected to left (L) and right (R) semi-infinite leads. The device area is described by the Hamiltonian $H_D$ and corresponding overlap matrix $S_D$. The conductance at zero temperature and in low bias limit can be obtained from the Landauer's theory for transport

$$G(E) = G_0 Tr\left(\Gamma_L \mathcal{G} \Gamma_R \mathcal{G}^\dagger\right), \qquad (1)$$

where $G$ is the conductance expressed within Fisher-Lee relationship [32], while $G_0 = 2e^2/h$ is the conductance quantum, $E$ is the energy of the incident charge carrier, and $Tr$ denotes the trace of corresponding operator; $\mathcal{G}$ is the retarded Green's function of the functionalized linear cubane-based chain, and matrices $\Gamma_{L,R} = i\left(\Sigma_{L,R} - \Sigma_{R,L}^\dagger\right)$ consider the coupling to the right or left leads through their self-energies $\Sigma_{L,R}$. The self-energies of the device-leads system can be described as

$$\begin{aligned}\Sigma_L &= \left[(E+i\eta)S_{LD} - H_{LD}\right]^\dagger \left\{(E+i\eta)S_L^0 - H_L^0 + \left((E+i\eta)S_L^{01} - H_L^{01}\right)^\dagger \bar{T}_L\right\}^{-1} \left[(E+i\eta)S_{LD} - H_{LD}\right]; \\ \Sigma_R &= \left[(E+i\eta)S_{DR} - H_{DR}\right]\left\{(E+i\eta)S_R^0 - H_R^0 + \left((E+i\eta)S_R^{01} - H_R^{01}\right)^\dagger T_R\right\}^{-1}\left[(E+i\eta)S_{DR} - H_{DR}\right]^\dagger,\end{aligned} \qquad (2)$$

where $\eta$ is an arbitrary small quantity; $H_{LD}$, $H_{DR}$, $S_{LD}$, $S_{DR}$ are the coupling Hamiltonians and corresponding overlap matrices describing connections device-left lead and device-right lead; $H_L^0$, $H_R^0$, $H_L^{01}$, $H_R^{01}$, $S_L^0$, $S_R^0$, $S_L^{01}$, $S_R^{01}$ describe the left/right lead unit cell and coupling between the neighboring unit cells of the lead; $\bar{T}_L$ and $T_R$ are the lead transfer matrices that can be easily computed from $H_L^0$, $H_L^{01}$, $S_L^0$, $S_L^{01}$ and $H_R^0$, $H_R^{01}$, $S_R^0$, $S_R^{01}$ via an iterative procedure (see Refs. 35-38 for details). Note that in our case left and right leads as well as device region have the same molecular structure. Therefore $H_L^0 \equiv H_R^0 \equiv H_D$, $S_L^0 \equiv S_R^0 \equiv S_D$, and $H_L^{01} \equiv H_R^{01} \equiv H_{LD} \equiv H_{DR}$, $S_L^{01} \equiv S_R^{01} \equiv S_{LD} \equiv S_{DR}$. The retarded Green's function is given by the expression

$$\mathcal{G}(E) = \left[(E+i\eta)S_D - H_D - \Sigma_L - \Sigma_R\right]^{-1}. \qquad (3)$$

The electronic density of states (DOS) is given by

$$\mathrm{DOS}(E) = \left(-\frac{1}{2\pi}\right) Tr\left\{\left(\mathcal{G} - \mathcal{G}^\dagger\right) S_\mathrm{D}\right\}. \qquad (4)$$

The electronic Hamiltonians are calculated from overlap matrix elements using extended Hückel approximation [39] with Anderson distant dependence for Wolfsberg-Helmholz parameter [40]. The overlap integrals are calculated using standard Slater-Koster-Roothaan procedure [41-42]. Therefore, the analytical expressions for the overlap integrals greatly simplify the calculation of Hamiltonian. Fact mentioned above make this approach very efficient for recursive Green's function method as described in Ref. 37.

## 3. Results and discussions

At first, we construct CL-20 chains consisting from two to eight fragments interconnected via the $CH_2$ molecular bridges (Fig. 1c). Earlier we obtained that dimers containing carbon-based molecular bridges are more thermodynamically stable than those, in which CL-20 fragments are bound through the nitrogen-based ones, for example, N–$NO_2$ [15]. Additional geometry optimization in the frame of this work demonstrate the instability of CL-20 chains of any length constructed from fragments connected via the $NO_2$ and $NH_2$ functional groups. Thus, only carbon-based molecular bridges conserve the identity of the elementary CL-20 units. Next, we calculated binding energies $E_b$ and HOMO-LUMO gaps $\Delta_{HL}$ of the chains constructed. The binding energy $E_b$ of the chain per atom is determined by the equation

$$E_b\left[\frac{\mathrm{eV}}{\mathrm{atom}}\right] = \frac{1}{N_{at}}\left\{kE(\mathrm{H}) + lE(\mathrm{C}) + mE(\mathrm{N}) + nE(\mathrm{O}) - E_{tot}(\text{CL-20 chain})\right\}, \qquad (5)$$

where $N_{at} = k + l + m + n$ is the total number of atoms in the chain, $E_{tot}(\text{CL-20 chain})$ is the total CL-20 chain energy, $E(\mathrm{H})$, $E(\mathrm{C})$, $E(\mathrm{N})$, and $E(\mathrm{O})$ are the energies of the isolated hydrogen, carbon, nitrogen, and oxygen atoms, respectively. The chain with higher binding energy (lower potential energy) is more thermodynamically stable and vice versa. HOMO-LUMO gap is defined as energy gap between the highest occupied molecular orbital and the lowest unoccupied molecular orbital. The binding energies $E_b$ obtained for finite CL-20 chains consisting from two to eight fragments are presented on Fig. 3.

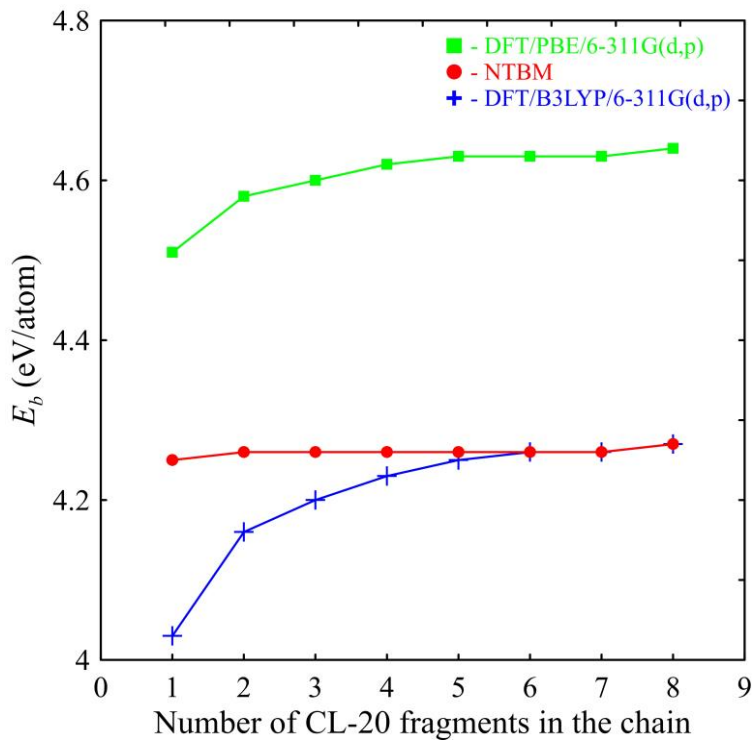

**Figure 3.** Binding energies versus the number of CL-20 fragments in the chain obtained at the DFT/B3LYP/6-311G(d,p) (crosses), DFT/PBE/6-311G(d,p) (squares), and NTBM (circles) levels of theory

As evident from the Fig. 3, as the number of CL-20 fragments increases, binding energy becomes larger. The increase of binding energy indicates that CL-20 chains become more thermodynamically stable as their effective lengths increase. Maybe the high thermodynamic stability of "long" CL-20 chains is associated with the decreasing of boundary induced structural distortions associated with the nitrogen containing functional groups as the number of CL-20 fragments increases. Thus, the formation of 2D-crystals or bulk covalent CL-20 solids may be energetically favorable, and such structures may possess much higher kinetic stability comparing to the CL-20 molecular crystals [43]. The HOMO-LUMO gaps $\Delta_{HL}$ calculated for finite CL-20 chains consisting from two to eight fragments monotonically decreases with the chain length (see Fig. 4). Such behavior is typical for various low-dimensional nanostructures, for example, carbon [44] or boron nitride [45] nanotubes, carbon polyprismanes [46], and graphene nanoribbons [47].

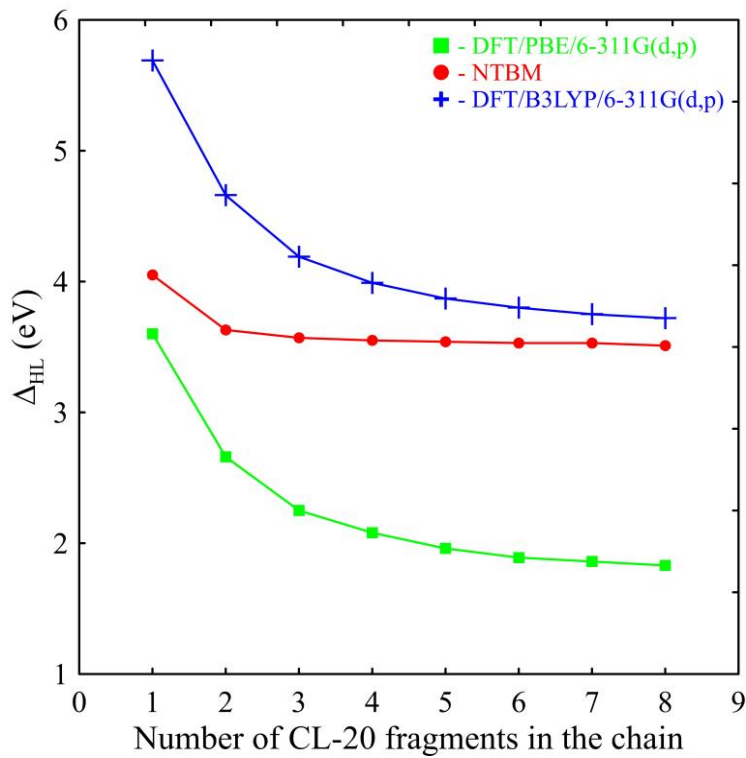

**Figure 4.** HOMO-LUMO gaps versus the number of CL-20 fragments in the chain obtained at the DFT/B3LYP/6-311G(d,p) (crosses), DFT/PBE/6-311G(d,p) (squares), and NTBM (circles) levels of theory

Note that HOMO-LUMO gaps obtained for "long" eight-fragment CL-20 chains is comparable with the upper limit of characteristic semiconducting value or, in other words, with the corresponding band gaps for the wide-bandgap semiconductors, such as GaP (2.3 eV), GaN (3.4 eV), and ZnSe (2.7 eV) [48]. In general data obtained using NTBM model are in good agreement with the DFT results, and are closer to B3LYP functional than to PBE. While PBE usually underestimates the HOMO-LUMO gap [49], B3LYP is much closer to the experimental data or slightly overestimates it [49-50]. So, in this case NTBM is a good choice to calculate the electronic characteristics of CL-20 chains, and further we use it to explore the conduction regime in the infinite CL-20 chains by calculating their conductance.

The CL-20 chain conductance as well as density of electronic states (DOS) is illustrated in Fig. 5.

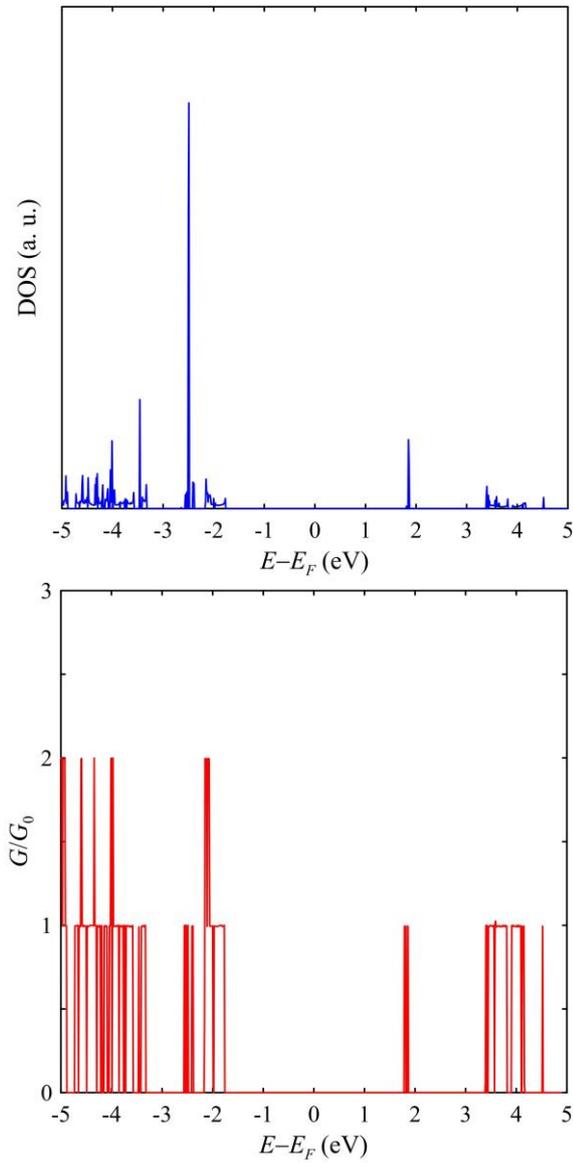

**Figure 5.** Density of electronic states (blue) and conductance (red, in units of $G_0=2e^2/h$) of the infinite CL-20 chain with the $CH_2$ molecular bridges

It turns out that the CL-20 chain constructed using the $CH_2$ molecular bridges is a wide-bandgap semiconductor. Note that this result is in good agreement with the HOMO-LUMO gaps of corresponding finite eight-fragment sample. Therefore, in the bulk limit for the covalent CL-20 crystal, it may be possible to transport electrons, but additional doping with another molecular bridges or mechanical stresses may be needed for electronic properties tuning that will be useful for nanoelectronic applications. Additional doping may lead to the opening of new conduction channels, energy gap may become narrower since the dopants states will be located near the Fermi level, and therefore, the conductivity will become possible.

We earlier obtained the similar result for the functionalized cubane-based chains [26], and we plan to verify this hypothesis for CL-20 covalently bounded systems in future.

## 4. Conclusions

We have constructed CL-20 chains using $CH_2$ molecular bridges for the covalent bonding between the isolated CL-20 fragments. It is shown that such type of conjunction makes the CL-20 based nanosystems high thermodynamically stable. For the one-dimensional chains, their stability increases with the efficient length growth. Therefore, the formation of CL-20 two-dimensional layers or crystalline covalent solids seems to be energetically favorable. On the other hand, numerical calculation of CL-20 chains electronic properties reveals that they are wide-bandgap semiconductors. Thus, the use of pure CL-20 chains with $CH_2$ molecular bridges in nanoelectronic applications seems to be problematic. However, this drawback may be solved introducing the additional molecular bridges into the chain or mechanical stresses application. We hope that this study will stimulate the subsequent theoretical and experimental research of CL-20 covalent crystals.

**Author contributions**

Konstantin Katin and Mikhail Maslov contributed equally to this paper. They performed theoretical calculations and wrote the manuscript.

**Acknowledgements**

The reported study was financially supported by Grant of the President of the Russian Federation, Grant No. MK-7410.2016.2.